\documentclass[10pt,twoside,journal]{IEEEtran}
\usepackage{graphics}
\usepackage{amsmath}
\usepackage{latexsym}
\usepackage{epstopdf}

\newcommand{\ket}[1]{\vert #1 \rangle}
\newcommand{\dyadic}[1]{{#1}
\setbox0=\hbox{$\scriptstyle\leftrightarrow$}
   \setbox2=\hbox{$#1$}
   \dimen0=.5\wd0 \advance\dimen0 by-.5\wd2
   \advance\dimen0 by-\wd0
   \kern\dimen0
{^{\hbox{$\scriptstyle\leftrightarrow$}}}}

\begin{document}

\title{Broadband Rydberg Atom-Based Electric-Field Probe:
From Self-Calibrated Measurements to Sub-Wavelength Imaging}

\author{~Christopher~L.~Holloway,~\IEEEmembership{Fellow,~IEEE,}
        Josh A. Gordon,  Steven Jefferts, Andrew Schwarzkopf, David A. Anderson, Stephanie A. Miller, Nithiwadee Thaicharoen, and Georg Raithel
        \thanks{Manuscript received \today.}\thanks{C.L. Holloway, J.A. Gordon, and S. Jefferts, are with the National
Institute of Standards and Technology (NIST), U.S. Department of Commerce, Boulder Laboratories,
Boulder,~CO~80305. A. Schwarzkopf, D.A. Anderson, S.A. Miller, N. Thaicharoen, and G. Raithel are with the Department of Physics, University of Michigan, Ann Arbor, MI 48109.  This work was partially supported by DARPA's QuASAR program. Publication of the U.S. government, not subject to U.S. copyright.}}

\markboth{ }{Holloway, et al., Rydberg Atom-Based Electric-Field Probe: Self-Calibrated Measurements to Sub-Wavelength Imaging}

\maketitle

\begin{abstract}
We discuss a fundamentally new approach for the measurement of electric (E) fields that will lead to the development of a broadband, direct SI-traceable, compact, self-calibrating E-field probe (sensor). This approach is based on the interaction of radio frequency (RF) fields with alkali atoms excited to Rydberg states.  The RF field causes an energy splitting of the Rydberg states via the Autler-Townes effect and we detect the splitting via electromagnetically induced transparency (EIT).  In effect, alkali atoms placed in a vapor cell act like an RF-to-optical transducer, converting an RF E-field strength measurement to an optical frequency measurement. We demonstrate the broadband nature of this approach by showing that one small vapor cell can be used to measure E-field strengths over a wide range of frequencies: 1~GHz to 500~GHz. The technique is validated by comparing experimental data to both numerical simulations and far-field calculations for various frequencies. We also discuss various applications, including: a direct traceable measurement, the ability to measure both weak and strong field strengths, compact form factors of the probe, and sub-wavelength imaging and field mapping.

\vspace{2mm}
\end{abstract}

\section{Introduction}

\begin{figure}[!t]
\centering
\scalebox{.38}{\includegraphics*{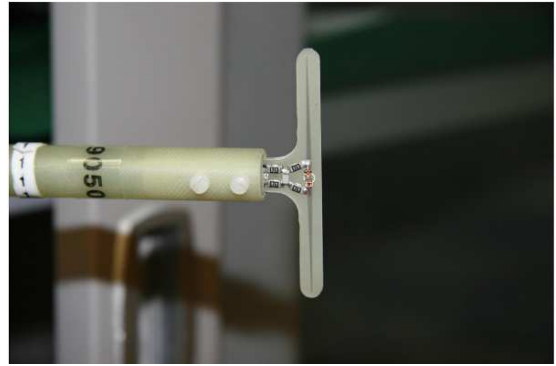}}
\caption{Common type of dipole probe.}
\label{dipoleprobe}
\end{figure}

Calibrating an electric ($E$) field probe and/or measuring an $E$-field can be challenging, and is somewhat of a chicken-or-egg dilemma. In that to calibrate a probe, one must place the probe (sensor) in a ``known'' field.
However, to know the field we need a calibrated probe. There are various types of probes used.  One example is the probe shown in Fig.~\ref{dipoleprobe}, which consists of a diode placed across a dipole antenna. The output of the diode is connected to a DC volt-meter via a high-impedance line (on the order of 10,000~k$\Omega$/m) \cite{dipole}.  When placed in an E-field, the diode rectifies the electromagnetic (EM) field and the DC voltage is recorded. This DC voltage increases or decreases with an increasing and decreasing EM field strength.  To use this dipole probe, it first must be calibrated, which involves placing the probe in a known (calculated) field.

The probes discussed above are typically used at frequencies well below the resonant frequency of the dipole, that is, these probes are electrically small.  This is necessary to minimize perturbation of the measured field by the probe and to avoid wide variations in frequency response near the resonant frequency of the dipole.  The impedance of the non-resonant dipole elements along with the characteristics of the voltage detector (diode and rectifying circuit) can result in a relatively constant response over a wide frequency range.  This constant response is limited by the resonant frequency of the dipole at the upper end and the diminished response due to the small physical length at the lower frequencies.  The 5-cm length probe shown in Fig.~\ref{dipoleprobe} is useful as a probe up to about 2~GHz with calibration.  Probes for higher frequencies (up to 40~GHz) have been constructed using much smaller dipole elements (8~mm or less) with tapered resistive dipole elements that suppress the resonance \cite{dipole}.  The combination of very small dipole antennas, and diode detector circuits connected to resistive lines and high-impedance voltmeters, requires substantial $E$-field strength for reliable measurements.  While fields can be detected in the range of 100~mV/m the amplitude uncertainties can be large.  The useful sensitivity (minimum field strength) is typically on the order of 500-1000~mV/m \cite{dipole2} and \cite{dipole3}.   While the probes can in principle be electrically small (with the caveat that the sensitivity decreases as the probe size becomes smaller \cite{dipole3}), the limiting factor on the overall size of the probe is due to both the electronics in the probe head and the  size of dipole antenna required to drive the current across the diode.

While this type of probe has been used for over 40 years, it does have the following limitations: (1) it needs be to calibrated, (2) the sensitivity of the probe is governed by the dipole length, (3) the metal in the probe perturbs the field being measured, and (4) the sensitivity of the probe is limited to a minimal detectable field strength of 100~mV/m.  Besides the dipole probe shown in Fig. \ref{dipoleprobe}, there are other types of probes. For example there are probes based on non-linear materials (e.g., lithium niobate crystals), where the phase of an optical signal propagating through this material changes when immersed in an E-field.  While these types of probes can gain about one order of magnitude in sensitivity ($\sim 10$~mV/m), they still require calibration and will perturb the field being measured.

In order to calibrate a probe, the probe is placed in a ``known'' field.  The most common way the generate a known field is to perform a measurement in an anechoic chamber (AC) or other type of test facility. For an AC configuration, the probe is placed at a known distance (say $x$, typically 1~to~3~m) from an antenna (typically a horn antenna or open-ended waveguide). With Maxwell's equations and the dimensions of the horn antenna, the field strength at the distance $x$ is calculated for a given input power to the horn antenna.  The probe is placed in this ``known'' field and the output of the probe is recorded.  Due to the uncertainties in this approach, the ``known'' field is only typically known to within 5~$\%$ (or 0.5~dB) \cite{nistnote}.

Thus, common E-field probes in use today have many shortcomings: they are not very sensitive, may perturb the field during the measurements, may be relatively large, and require a calibration. The calibration procedure relies on a field value that is known to within only 5~$\%$.  One promising approach to remedy these problems is by using an E-field probe based on room-temperature Rydberg atoms. In this technique, alkali atoms placed in a vapor cell (a glass cell with atomic vapor inside, see Fig. \ref{vaporcell}(a)) are excited optically to Rydberg states and the applied radio frequency (RF) field alters the resonant state of the atoms. (Throughout this paper, the term ``RF'' is used to cover the conventional RF, microwave, millimeter wave, and sub-terahertz frequency ranges.) This approach exploits the sensitivity of the high-lying Rydberg states to RF radiation.  This sensitivity is reflected by the large transition matrix elements ($\wp$, on the order of $10^3$ to $10^4 e a_0$, where $e$ is the electric charge and $a_0$ is the Bohr radius) for RF transitions between Rydberg states. We measure an Autler-Townes splitting \cite{autler} of Rydberg energy levels in these atoms due the applied RF field. This splitting is easily measured with electromagnetically induced transparency (EIT) \cite{EIT}-\cite{EIT2} and is directly related to the applied E-field strength, Planck's constant $\hbar$, and $\wp$. A measurement of the splitting gives a measure of the field strength.  The high accuracy in this approach is because the EIT technique reduces an amplitude measurement (the desired quantity) to a frequency measurement (a measurement that can be performed very accurately). It is possible to excite the atom to a wide range of atomic states (a state that can interact with the applied RF E-field). As such, with one vapor cell, accurate measurements of a RF E-field strength over a frequency range from 1~GHz to 500~GHz are possible.

There is a push from various international metrology laboratories (including the National Institute for Standards and Technology, NIST) to make all measurements traceable to SI-units and/or traceable to fundamental physical constants.  While, a large number of various measurements are SI traceable, to date, all methods to make an E-field measurement that is SI traceable requires a complex traceability path.
The technique discussed in this paper provides a much more direct traceability path.

This new approach for E-field measurements has the following benefits: (1) it yields the field strength in SI units from a frequency measurement, fundamental constants, and known atomic parameters, (2) it is self-calibrating due to the invariance of the atomic parameters, (3) it will provide RF E-field measurements independent of current techniques, (4) since no metal is present in the probe, the probe will cause minimal perturbation of the field during the  measurement, (5) it will measure both very weak and very strong fields over a large range of frequencies (field strengths as low as 0.8~mV/m have been measured, and below 0.01~mV/m may be possible \cite{jim}), and (6) it allow for the construction of small, compact probes (optical fiber and chip-scale probes).  Possible applications for this probe are numerous, ranging from biomedical to sub-wavelength imaging.

Atomic measurement standards have been used for a number of years for a wide array of measurements, most notable are time, frequency, and length. However there are just a few publications on the use of the atom for $E$-field measurements \cite{jim}-\cite{carter}.  There is also work on the measurement of DC $E$-fields \cite{neukammer}-\cite{bason}. In this paper we will present the underlying physics of the EIT technique in a manner familiar to the EM community and demonstrate how it can be used in the development of a broadband probe covering a frequency range of 1~GHz to 500~GHz. We will also present calculations of the required $\wp$ for a specific set of atomic states for a range of RF transitions which are needed to determine the E-field from this measured splitting. The technique is validated by the agreement of experimental data with both numerical simulations and far-field calculations for various frequencies. We discuss the uncertainties in this approach and discuss various current and potential applications.


\section{EIT and Autler-Townes Splitting}

\begin{figure}[!t]
\centering
\scalebox{.9}{\includegraphics*{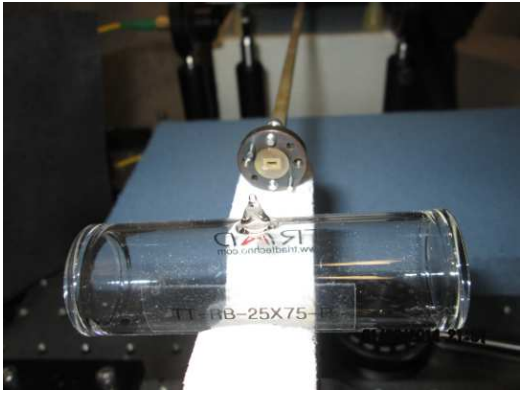}}\\
\scriptsize{\centerline{(a)}}
\vspace{2mm}\scalebox{.40}{\includegraphics*{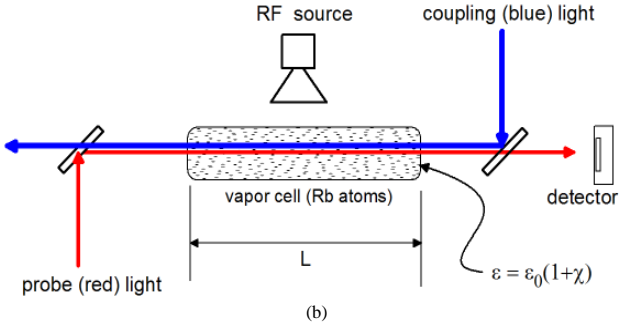}}
\scriptsize{\centerline{(b)}}
\caption{Illustration of a vapor cell and measurement setup: (a) Cylinder vapor cell, (b) Vapor cell setup for measuring EIT, with counter-propagating probe (red) and coupling (blue) beams. The RF is applied transverse to the optical beam propagation in the vapor cell.}
\label{vaporcell}
\end{figure}

In this section we present the basic concept of the measurement approach, which uses a vapor of alkali atoms, contained in a cell (see Fig.~\ref{vaporcell}), as the active medium of the probe. We choose atomic species that have a sufficiently high vapor pressure at room temperature.
In this paper, we concentrate on rubidium-85 ($^{85}$Rb) atoms, but other alkali atoms could be used. A diagram of the measurement setup with the $^{85}$Rb vapor cell is shown in Fig.~{\ref{vaporcell}}(b).

\begin{figure}[!t]
\centering
\scalebox{.25}{\includegraphics*{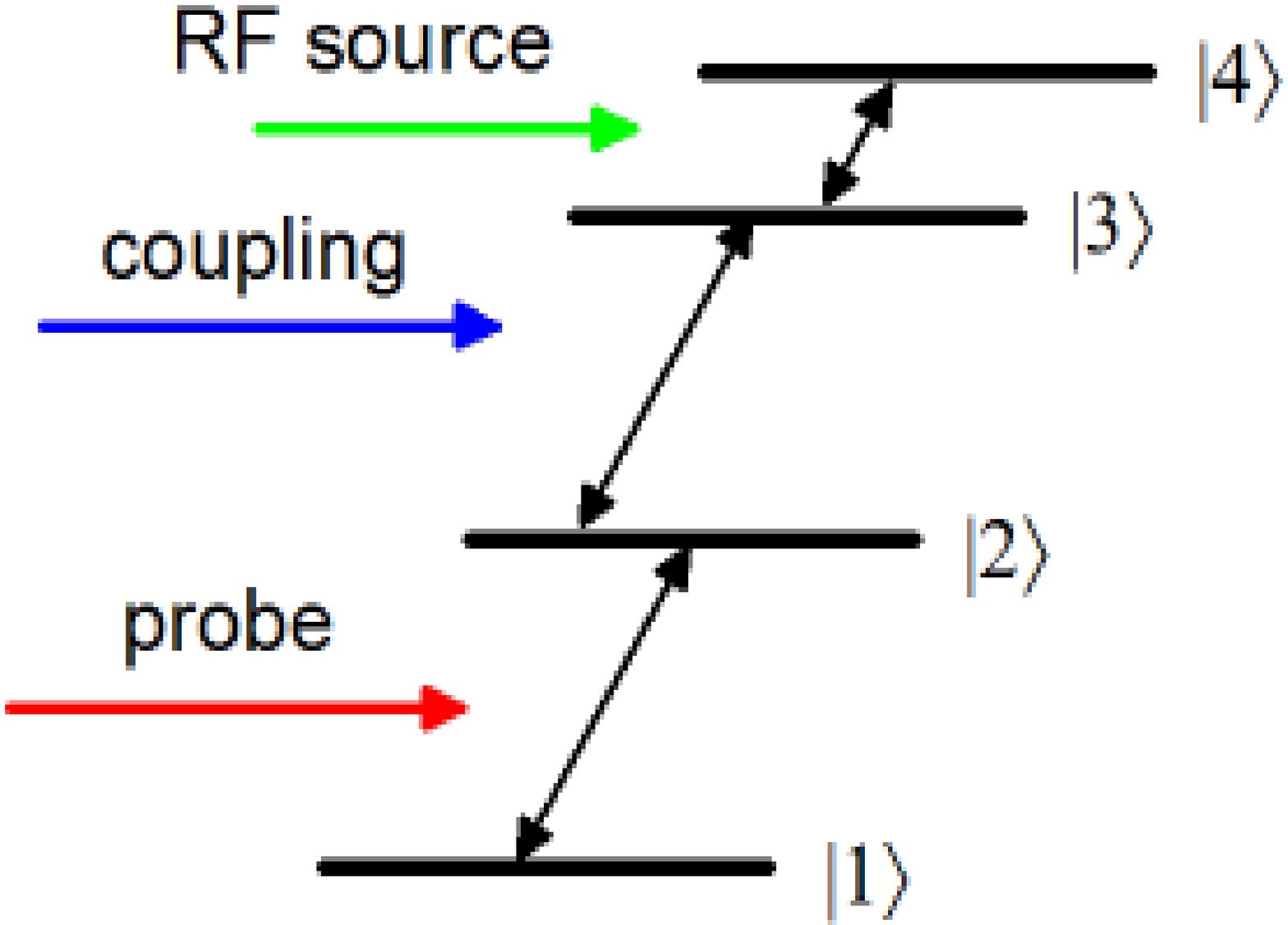}}
\caption{A four-level atomic system. Here, the atomic states are labeled as $\ket{1}$, $\ket{2}$, $\ket{3}$, and $\ket{4}$.}
\label{4level}
\end{figure}

The relevant atomic states for this approach are in the four-level atomic system depicted in Fig.~\ref{4level}.  When the frequency of the light (probe laser) matches the $\ket{1}$ to $\ket{2}$ atomic resonance (using $^{85}$Rb for our application, this corresponds to a 780~nm or ``red'' laser), the atoms scatter light from the incident beam and reduce the transmitted light intensity measured on the detector.  If a second strong (``coupling") light field is applied resonant with the $\ket{2}$ to $\ket{3}$ transition (using $^{85}$Rb for our application, this corresponds to a 480~nm or ``blue'' laser), the $\ket{2}$ and $\ket{3}$ states are mixed to form dressed state pairs which are close in energy. The excitation amplitudes from $\ket{1}$ to each of these two dressed states then have opposite signs, leading to destructive quantum interference of these excitation pathways. As such, a transparency window is opened for the probe (``red'') light: probe light transmission is increased. This is the phenomenon known as EIT \cite{EIT}.

If the atomic states $\ket{3}$ and $\ket{4}$ are chosen appropriately, an applied RF field will couple states $\ket{3}$ and $\ket{4}$. A third dressed state is then introduced between the two involved in EIT which leads to constructive interference in the probe absorption.  This splits the EIT resonance in two, and for resonant driving fields the new transmission maxima are split by the Rabi frequency $\Omega_{RF}$ (defined in detail below) of the $\ket{3}$-$\ket{4}$ transition \cite{Lukin1999} and \cite{Dutta2007}. This is known as Autler-Townes splitting  \cite{tony} of the EIT signal, which is related to the applied field and allows for a measurement of the E-field strength.

In order to measure the field strength (or amplitude) for different frequencies, different states $\ket{3}$ and $\ket{4}$ can be chosen.  State $\ket{3}$ is selected by tuning the wavelength of the coupling laser, and the $\ket{4}$ state is selected via the RF source.  In doing this, a large range of atomic transitions can be selected, allowing measurements of RF fields over a correspondingly wide selection of frequencies. In essence, the atoms act as highly-tunable, resonant, frequency selective  RF detectors.  This is a significant benefit of using Rydberg atoms as field probes.
The wide range of states $\ket{3}$ selectable by the coupling laser, translates to the broadband nature of the probe,
which allows RF measurements ranging from 1~GHz to 500~GHz.

There are various ways to analytically describe and model this technique.  One convenient approach from an EM perspective is to model the atoms in the vapor cell as an effective medium in which the effective permittivity of the medium for the probe laser propagating through the cell is given by
\begin{equation}
\epsilon=\epsilon_0\left(1+\chi\right) \,\,\, ,
\end{equation}
where $\chi$ is the susceptibility of the medium for the probe laser. Using the results in \cite{sandhya} and \cite{meystre},
the susceptibility is given by
\begin{equation}
\textstyle{
\chi=\frac{jN|\wp_{p}|\Omega_p}{|E_p|\epsilon_0}
\frac{\left(\Omega_{RF}\right)^2+4\,D_{13}D_{14}}
{D_{12}\left(\Omega_{RF}\right)^2
+D_{14}\left(\Omega_{c}\right)^2
+4\,D_{12}D_{13}D_{14} }
} \,\,\, ,
\label{chieq}
\end{equation}
where
\begin{equation}
D_{1i}=\gamma_{1i}-j\Delta_p \,\,\, ,
\end{equation}
$N$ is the atom density in the cell, and the subscripts 12, 13, and 14, correspond to the transitions of from the ``i'' to ``1'' state (i.e., $\ket{i}$ to $\ket{1}$, see the labels in Fig.~\ref{4level}). The parameter $\gamma_{1i}$ is the decay rate for the various states (where ``i'' is 2-4,), $\Delta_p$ is the de-tuning of the probe laser (defined as $\Delta_p=\omega_o-\omega_p$, where $\omega_o$ is the on-resonance angular frequency of states $\ket{1}$ to $\ket{2}$ and $\omega_p$ is the angular frequency of the probe laser). The quantities $\Omega_{p,c,RF}$ are the Rabi frequencies for the different transitions and are given by
\begin{equation}
\Omega_{p, c, RF}=|E_{p, c, RF}|\frac{\wp_{p, c, RF}}{\hbar} \,\,\, ,
\label{rabi}
\end{equation}
where $\hbar$ is Planck's constant, $|E_{p, c, RF}|$ are the E-field of the probe laser, the coupling laser, and the RF source, respectively. Finally, $\wp_p$, $\wp_c$ and $\wp_{RF}$ are the atomic dipole moments corresponding to the probe, coupling, and RF transitions. We should add that in general, the parameters $D_{13}$ and $D_{14}$ would also be a function of $\Delta_c$ and $\Delta_{RF}$ (the de-tuning of the coupling laser and RF source), see \cite{sandhya}.  Here, we take the coupling laser frequency and RF to be resonant with their respective transitions (or $\Delta_c=\Delta_{RF}=0$) and only consider a detuning of the probe from the $\ket{1}$ to $\ket{2}$ transition.

Using this susceptibility, the magnitude of the transmission coefficient ($T$) of the probe laser propagating through the cell can be approximated by the following \cite{yariv}:
\begin{equation}
|T|=\exp\left(-\frac{2\pi L \,\,{\rm Im}\left[\chi\right]}{2\lambda_p}\right)\,\,\, ,
\end{equation}
where $L$ is the length of the cell and $\lambda_p$ is the wavelength of the probe laser.  The intensity of the probe beam measured on the detector
is given by
\begin{equation}
I=I_0 |T|^2 =I_0 \exp\left(-\frac{2\pi L \,\,{\rm Im}\left[\chi\right]}{\lambda_p}\right) \,\,\, ,
\label{intensity}
\end{equation}
where $I_0$ is the intensity of the probe beam at the input of the cell.

\begin{figure}[!t]
\centering
\scalebox{.61}{\includegraphics*{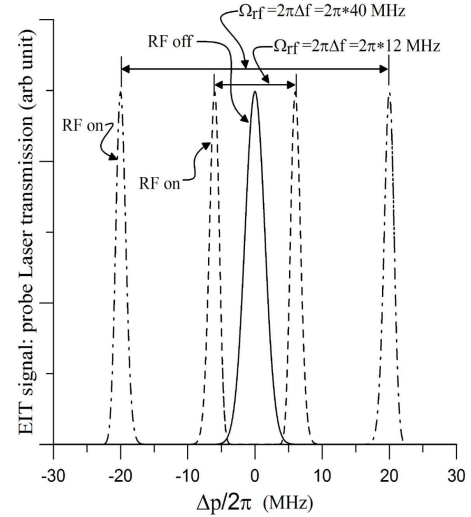}}
\centerline{\scriptsize{(a) EIT signal}}
\vspace{2mm}
\scalebox{.65}{\includegraphics*{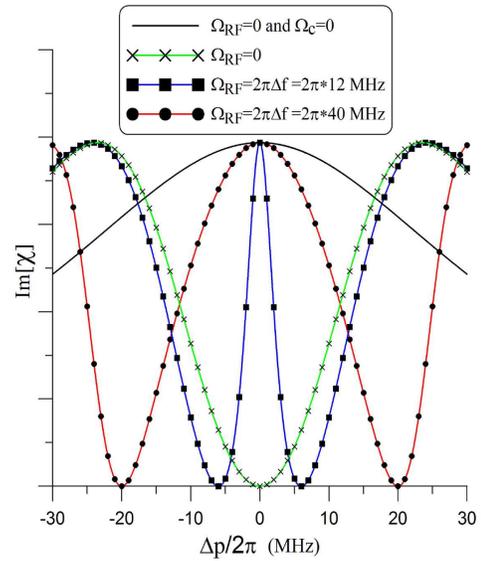}}
\centerline{\scriptsize{(b) ${\rm Im}[\chi]$}}
\caption{EIT signal and ${\rm Im}[\chi]$ for different $\Omega_{RF}$ as function of $\Delta_p$ for $\Omega_{c}=$40~MHz.}
\label{model}
\end{figure}

We use this model to understand the behavior of the EIT signal on the detector as a function of the applied RF field strength (i.e., $\Omega_{RF}$). Fig.~\ref{model}(a) illustrates the EIT signal (i.e., $|T|^2$) as a function of $\Delta_p$ for increasing $\Omega_{RF}$ [or increasing the applied RF-field strength $|E_{RF}|$ through (\ref{rabi})]. These results were obtained with $\gamma_{12}=2\pi\cdot$6.066~MHz, and $\gamma_{13}=\gamma_{14}=10^{-4}\cdot\gamma_{12}$ (which are typical values for the transitions discussed in this paper \cite{dline}). We see that as $\Omega_{RF}$ increases, the splitting between the two peaks of the EIT signal increases. To understand this from an effective material viewpoint, it is instructive to look at the behavior of ${\rm Im}[\chi]$ as a function of the applied RF field, see Fig.~\ref{model}(b).  Once the RF field is applied, we see that at two locations, ${\rm Im}[\chi]$ goes to zero, which corresponds to no absorption, and hence a large signal on the detector (this is the location of the transparency window which produces the EIT signal).  As the RF field strength is increased the separation in the zero locations increases, corresponding to an increased separation in the peaks of the EIT signal. For a reference, we have also plotted the case with no blue laser power (i.e., $\Omega_c=0$). For this case, ${\rm Im}[\chi]$ is maximum at $\Delta_p=0$, which implies maximum absorption and no EIT signal. On the other hand, when $\Omega_{c}\ne0$ and $\Omega_{RF}=0$ (the line marked with crosses), we see that ${\rm Im}[\chi]=0$ at $\Delta p=0$ and therefore, we have no absorption (the EIT signal with no splitting). Referring back to Fig.~\ref{model}(a), the observed splitting of the two peaks when $\Omega_{RF}\ne0$ is referred to as Autler-Townes splitting ($\Delta f_0$, measured in hertz), and is related to the Rabi frequency ($\Omega_{RF}=2\pi\Delta f_0$). In fact, it is possible to show that the separation between the two peaks in the probe spectrum is simply the Rabi frequency associated with the RF field transition \cite{berman}, or
\begin{equation}
peak-to-peak=\Delta f_0=\frac{\Omega_{RF}}{2\pi}\,\,\, .
\end{equation}
The Rabi frequency $\Omega_{RF}$ (or $\Delta f_0$) can be measured and the magnitude of the applied RF field can be determined by (\ref{rabi})
or
\begin{equation}
|E_{RF}|=\frac{\hbar}{\wp_{RF}}\Omega_{RF}=2\pi\frac{\hbar}{\wp_{RF}}\Delta f_0 \,\,\, .
\label{ecal2}
\end{equation}
Thus, to measure $|E_{RF}|$, $\Omega_{RF}$ (or $\Delta f_0$) is obtained from a frequency measurement (that can be measured very accurately), Planck's constant is known, and the only unknown is the dipole moment $\wp_{RF}$, which can be calculated using first principles as shown below.
We consider this type of measurement of the E-field strength a direct SI traceable measurement in that it is related to Planck's constant.
By using the interaction of atoms in a vapor cell with two applied lasers and a RF field, it is possible to reduce an RF E-field amplitude measurement to a frequency measurement (or simply measuring the frequency difference in the splitting of an EIT signal).

As a side note, the EIT technique has interesting possibilities from a metamaterials viewpoint. There is a great deal of attention in the metamaterial community on developing tunable materials that can be used for slow-wave and near-zero permittivity materials, see \cite{EIT}, \cite{nader1}, and \cite{nader2}. This EIT technique can in principle be used to tailor the real part of effective permittivity at the probe-laser wavelength via an RF source, i.e., for a given probe laser wavelength, an applied RF field can be used to obtain a desired value for $\epsilon$ (e.g., ${\rm Re}[\epsilon/\epsilon_0]<1$). This will be investigated in more detail in future publications.

\section{Broadband Nature of the Technique}

A conventional dipole antenna is tuned or optimized to a particular frequency by its physical size (or dipole length). This atom-based probe is somewhat analogous to the dipole antenna, in that, instead of varying a length dimension (the dipole length) we simply vary the wavelength of the coupling light to measure a desired RF transition. That is, the coupling laser is tuned to different wavelengths in order to measure the field strength at different frequencies. The precise wavelength of the coupling laser governs which atomic states can be used to measure this RF field strength, and the energy difference between these states determines the frequency of the RF field whose strength is measured. The significant benefit of this probe is that since the atom is a highly tunable resonator, we use that property to excite the atom to various states (with the coupling laser) such that it will respond to a wide range of frequencies, hence a broadband probe.

\begin{figure}[!t]
\centering
\scalebox{.20}{\includegraphics*{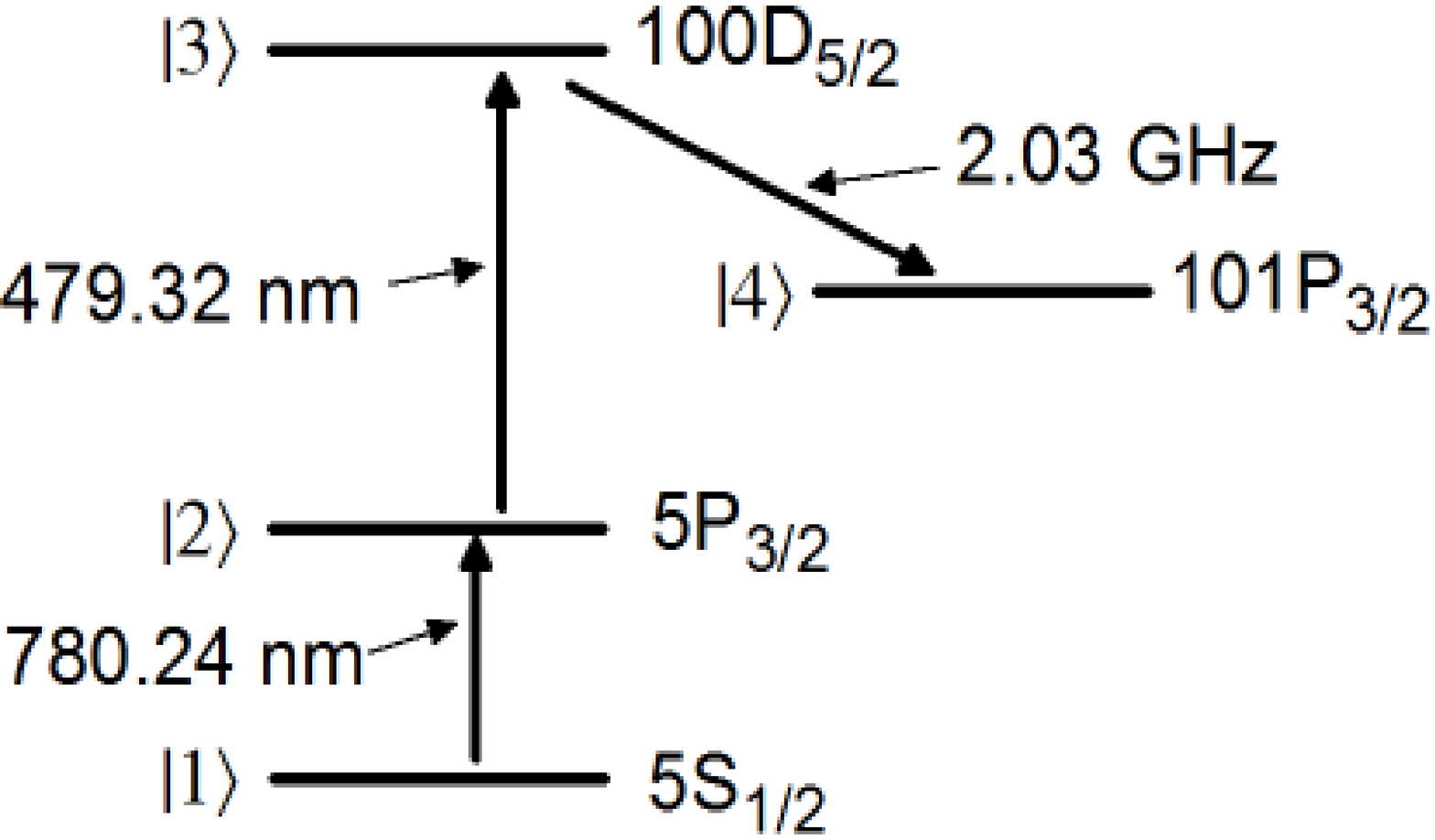}}
\hspace{3mm}
\scalebox{.20}{\includegraphics*{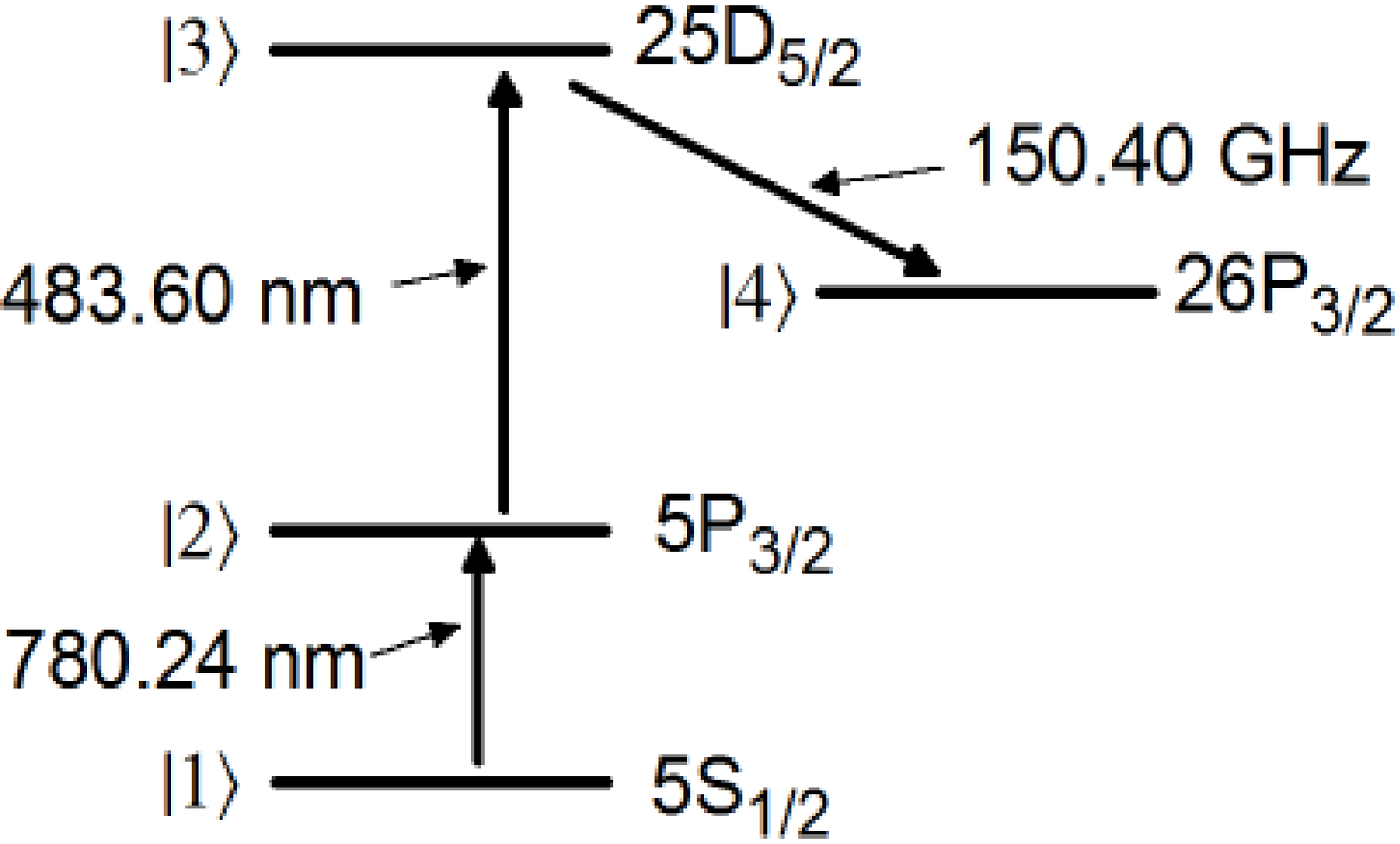}}\\
\scriptsize{\centerline{(a) \hspace{43mm} (b)}}
\caption{Four-level atomic system for $^{85}$Rb: (a) 2~GHz transition, and (b) 150~GHz transition. The ''S'', ``D'' and ``P'' indicate the angular momentum of the atomic state \cite{wolf}.}
\label{energylevel}
\end{figure}

In this paper, we concentrate on rubidium-85 ($^{85}$Rb) atoms; as such, the probe light is a 780~nm (``red'') laser and the $\ket{1}$ to $\ket{2}$ atomic resonance corresponds to the $5S_{1/2}-5P_{3/2}$ transition, see Fig.~\ref{energylevel}. To ensure that the $\ket{3}$ to $\ket{4}$ atomic resonance in $^{85}$Rb is an RF transition, the $\ket{2}$ and $\ket{3}$ transition will correspond to a $\sim$480~nm (``blue'') laser. Fig.~\ref{energylevel} depicts two four-level atomic systems that illustrate how two different blue wavelengths can result in two different RF transitions.  With the blue laser tuned to 479.32~nm, it is possible to measure an RF field strength at 2.03~GHz, while a blue laser at 483.60~nm will allow an RF measurement at 150.40~GHz.  This is the basic concept for the broadband probe.  With one vapor cell, a tunable red laser, and a tunable blue laser, it is possible to measure RF field strengths from 1~GHz to 500~GHz (we will illustrate this with experimental data below).

\begin{figure}
\centering
\scalebox{.4}{\includegraphics*{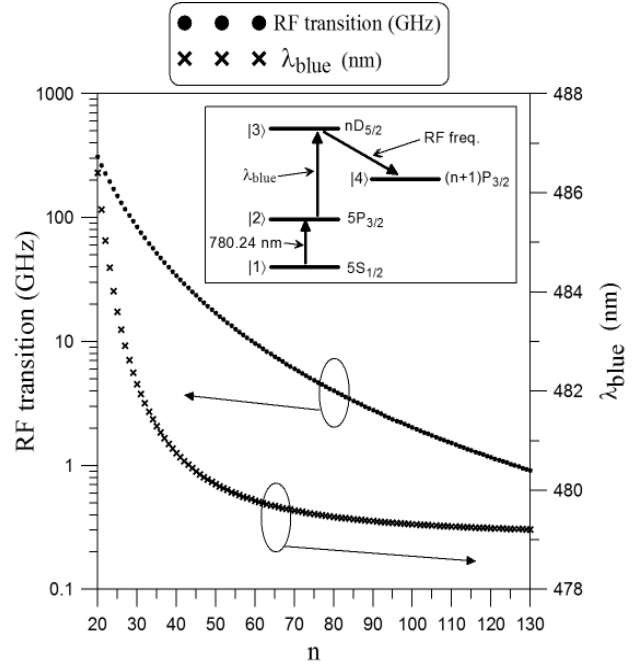}}
\caption{RF transition frequency and required blue laser wavelength for the $nD_{5/2}$-$(n+1)P_{3/2}$ transitions in $^{85}$Rb.}
\label{freq}
\end{figure}

An atom is ``rich'', in that the number of RF transitions that can be excited is numerous. In this paper it would be difficult to address all the possible atomic transitions (or states) that can be reached with an RF source.
From among the many transitions possible in 85Rb, we concentrate on transitions that occur between $nD_{5/2}-(n+1)P_{3/2}$ states, where $n$ is the principal quantum number. It is instructive to show the required blue wavelengths ($\lambda_{blue}$) needed to couple the $nD$ state from the red laser (or the $5P_{3/2}$ state). Also, it is instructive to show what frequencies can be measured for $nD_{5/2}-(n+1)P_{3/2}$ transitions.  These data is shown in Fig.~\ref{freq}.  From the figure we see that if $n$ ranges from 20 to 130, frequencies ranging from 300~GHz to 1~GHz can be measured. Also from this figure we see that to be able to measure this range of frequencies requires the blue laser to be tuned from 487~nm to 479~nm.   The results in Fig.~\ref{freq} were obtained using the Rydberg formula and the quantum defects for $^{85}$Rb \cite{gal}-\cite{gal3}.  This illustrates the broadband nature of this technique.  While the results in this figure only show the achievable frequencies for the $nD_{5/2}-(n+1)P_{3/2}$ transition, the measurement of other frequencies is possible with other atomic transitions.  This will be a topic of a future publication and is briefly discussed below.

\subsection{Atomic Dipole Moment}
\label{dipolesec}

After a measurement of the splitting, the one unknown needed for determining the E-field is the dipole moment $\wp_{RF}$ of the RF transition. The dipole moment is expressed as follows \cite{macclaff} and \cite{sobelman}
\begin{equation}
\wp_{RF} = e \, \vec{\epsilon} \cdot \int \Psi_3^{*} \,\,\vec{r} \,\,\Psi_4 dV \,\,\, ,
\label{wp}
\end{equation}
where $e$ is the electric charge, $\vec{\epsilon}$ is the polarization vector of the light, and $\Psi_3$ and $\Psi_4$ are the wavefunctions for the initial and final state for the RF transition. The wavefunctions are the solution of Schr\"{o}dinger's equation for each of these states.  The expression can be written as \cite{macclaff}
\begin{equation}
\wp_{RF}=e {\cal A}_{34} {\cal R}_{34}\,\,\, ,
\label{wp2a}
\end{equation}
where ${\cal A}_{34}$ is the angular part, and ${\cal R}_{34}$ is the radial part.

\begin{figure}[!t]
\centering
\scalebox{.35}{\includegraphics*{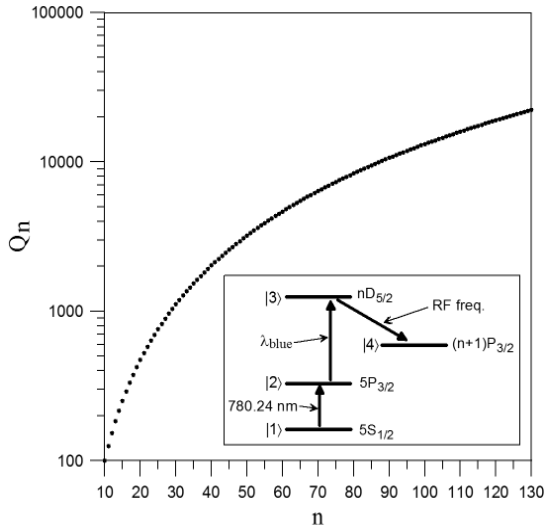}}
\caption{The normalized radial part of the dipole moment (${Q}_n={\cal R}/a_o$) for the $nD_{5/2}-(n+1)P_{3/2}$ transition in $^{85}$Rb.}
\label{rpart}
\end{figure}

Here again, we only address a small sub-set of the possible RF transitions. As such, we concentrate on RF transitions in $^{85}$Rb that occur between $nD_{5/2}-(n+1)P_{3/2}$ states. Furthermore, there are various polarization states (i.e., the polarization of the two laser fields relative to the polarization of the RF field) that affect which values of $A_{34}$ are relevant. In this paper we will only discuss and show results for linearly polarized laser fields and co-polarized RF fields.  (Investigating the relative difference of the polarizations of RF fields to the optical fields gives a possible way of measuring the full vector RF fields \cite{jim2}.)  For co-linear polarized optical and RF fields, and for $nD_{5/2}-(n+1)P_{3/2}$ states, it can be shown that (using expressions given in \cite{sobelman}) ${\cal A}_{34}=0.4899$.  The radial part ${\cal R}_{34}$ calculation requires first a numerical solution for the radial Schr\"{o}dinger's equation for the wavefunctions, and then a numerical evaluation of the radial integral. Using the method given in \cite{gal} (and the quantum defects in \cite{gal2, gal3}), we obtained numerical calculations of ${\cal R}_{34}$ for the $nD_{5/2}-(n+1)P_{3/2}$ states. These results are given in Fig.~\ref{rpart} for the scaled radial part (defined as $Q_n={\cal R}/a_0$, where $a_0$ is the Bohr radius). With those results, the required dipole moment is given by
\begin{equation}
\wp_{RF}=0.49 e a_0 Q_{n}\,\,\, .
\label{wp2}
\end{equation}

\begin{figure}
\centering
\scalebox{.35}{\includegraphics*{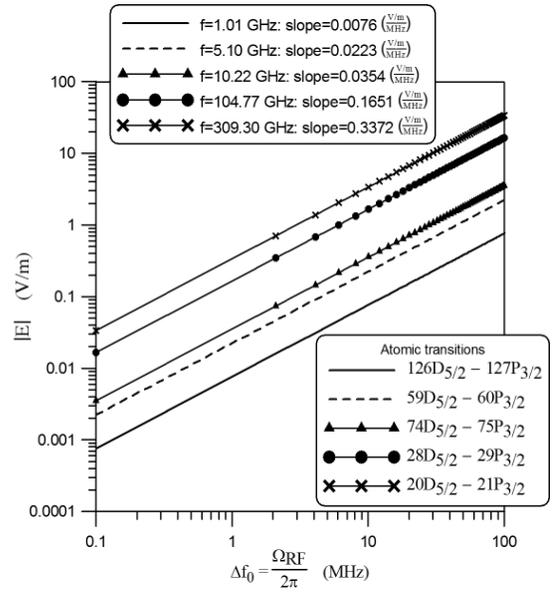}}
\caption{Family of curves for the sensitivity of the $E$-field measurement for $nD_{5/2}$-$(n+1)P_{3/2}$ transitions in $^{85}$Rb.}
\label{Mhzplot}
\end{figure}

From the plot given in Fig.~\ref{rpart}, we see that for $n$ greater than 90 (which corresponds to frequencies less than 3~GHz, see Fig.~\ref{freq}), the radial part of the dipole moments is greater than 10,000, and is larger than 1,000 for $n$ greater than 30 (frequencies less than 70 GHz). These large dipole moments reflect the sensitivity of this type of E-field measurement. With these dipole moments, we can generate a family of curves illustrating the slope of (or $2\pi{\hbar}/{\wp_{RF}}$) for the expression given in (\ref{ecal2}). The curves are shown in Fig.~\ref{Mhzplot}. These curves not only give one an indication of the slopes to expect in experimental data, but also give an indication of the type of sensitivity one can measure at the various frequencies. The sensitivity is a function of the frequency that is being measured. For example, at 1~GHz, a measured 1~MHz $\Delta f_0$ corresponds to a 7.6~mV/m; and at 300~GHz, a measured 1~MHz $\Delta f_0$ corresponds to 0.337~V/m.

\section{Experimental Setup}

\begin{figure}[!t]
\centering
\scalebox{2.56}{\includegraphics*{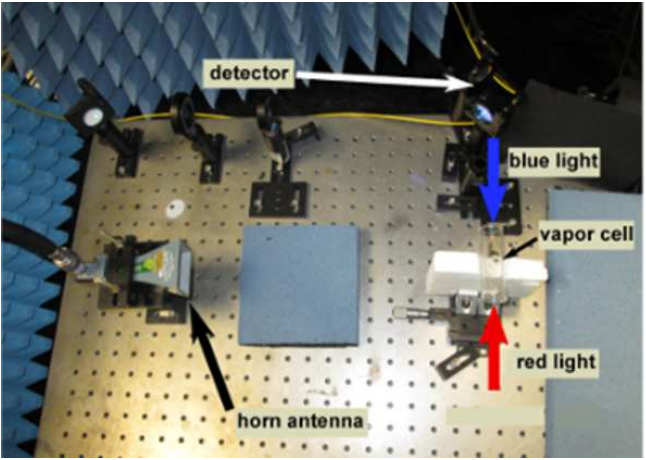}}
\caption{Experimental setup for $E$-field measurements with EIT.}
\label{exper}
\end{figure}

\begin{figure}[!t]
\centering
\scalebox{.28}{\includegraphics*{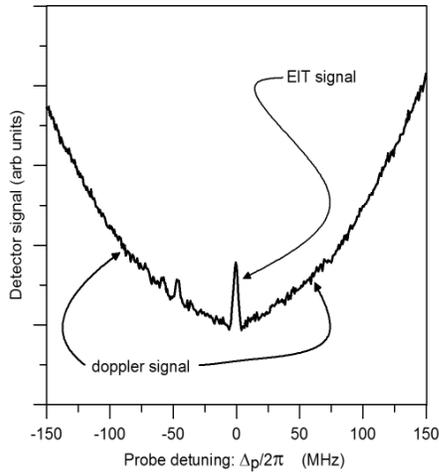}}
\caption{Probe transmission as a function of $\Delta_p$ for the three-level $5S_{1/2}-5P_{3/2}-50D$ EIT system.  The peak at $\Delta_p=0$ is the EIT signal we are investigating.}
\label{doppler}
\end{figure}

The experimental set-up used to demonstrate this approach is shown in Fig. \ref{exper}, which includes a vapor cell, a horn antenna (and a waveguide antenna is used for the higher frequency measurements), a lock-in amplifier,  a photo diode detector, a probe laser, and a coupling laser.  We use a cylindrical glass vapor cell of length 75~mm and diameter 25~mm containing ($^{85}$Rb) atoms.  The levels $\ket{1}$, $\ket{2}$, $\ket{3}$, and $\ket{4}$ correspond respectively to the $^{85}$Rb  $5S_{1/2}$ ground state,  $5P_{3/2}$ excited state, and two Rydberg states.  The probe
is a 780~nm laser which is scanned across the $5S_{1/2}$ -- $5P_{3/2}$ transition.  The probe beam is focused to a full-width at half maximum (FWHM) of 80~$\mu$m, with a power of order 100~nW to keep the intensity below the saturation intensity of the transition. Figure~\ref{doppler} shows a typical transmission signal as a function of relative probe detuning $\Delta_p$. The global shape of the curve is the Doppler absorption spectrum of $^{85}$Rb at room temperature. To produce an EIT signal, we apply a counter-propagating coupling laser (wavelength $\lambda_c \approx 480$~nm, ``blue") with a power of 22~mW, focused to a FWHM of 100~$\mu$m. As an example, tuning the coupling laser near the $5P_{3/2}$ -- $50D$ Rydberg transition results in distinct EIT transmission peaks as seen in the figure. Here we concentrate on the strongest peak at $\Delta_p=0$, labeled as ''EIT signal''. The other smaller peaks in this figure correspond to hyperfine sublevels and are discussed in \cite{imaging}.

\begin{figure}
\centering
\scalebox{.28}{\includegraphics*{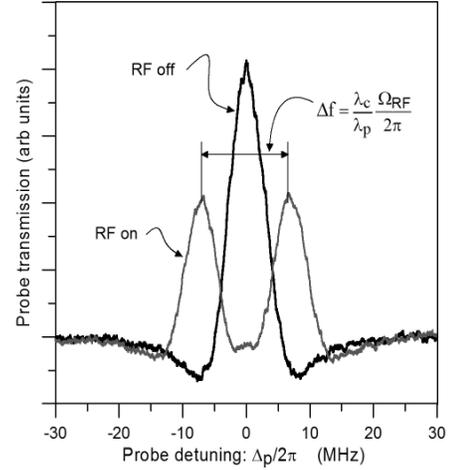}}
\caption{Black curve: EIT-signal as a function of $\Delta_p$ for the EIT system $5S_{1/2}-5P_{3/2} - 28D_{5/2}$. Gray curve: The $28D_{5/2}$ level is coupled to the $29P_{3/2}$ level by a 104.77~GHz RF field.}
\label{eitsignal}
\end{figure}

In order to improve the signal-to-noise ratio, we use heterodyne detection.  We modulate the blue laser amplitude with a 30~kHz square wave and detect any resulting modulation of the probe transmission with a lock-in amplifier.
This removes the Doppler background and isolates the EIT signal as shown in the black curve of Fig.~\ref{eitsignal}.
Here we tune the coupling laser near the $5P_{3/2}$ -- $28D_{5/2}$ transition (``blue'' with $\lambda_c \approx 482.63$~nm).  Application of a RF field at 104.77~GHz to couple states $28D_{5/2}$ and $29P_{3/2}$ splits the EIT peak as shown in the gray curve.

Differential Doppler shifts between the probe and coupling beams alter the frequency separations between EIT peaks in the probe transmission spectrum. Splittings of $5P_{3/2}$ hyperfine states are scaled by $1-\lambda_c/\lambda_p$, while splittings of Rydberg states are scaled by $\lambda_c/\lambda_p$ \cite{EIT_Adams}.  The latter factor is relevant to measurements of RF-induced splittings of EIT peaks and therefore (\ref{ecal2}) is modified.  We measure the frequency splitting of the EIT peaks in the probe spectrum, $\Delta f$, and determine the E-field amplitude by
\begin{equation}
	|E_{RF}| = 2 \pi \frac{\hbar}{\wp_{RF}} \frac{\lambda_p}{\lambda_c} \Delta f \quad.
	\label{mage2}
\end{equation}
Note here the use of the Doppler scaling factor, not present in (\ref{ecal2}) for stationary atoms.


\section{Experimental Results}

Using the experimental setup discussed above, we illustrate the measurement of E-field strengths for various frequencies.  Here we report on five of these: 15.59~GHz, 17.04~GHz, 18.65~GHz, 68.64~GHz, and 104.77~GHz. For the 15.59~GHz, 17.04~GHz and 18.64~GHz measurements a  horn antenna is used, and two different open-ended waveguides are used the 68.64~GHz and 104.77~GHz measurements.

\begin{figure}[!t]
\centering
\scalebox{.30}{\includegraphics*{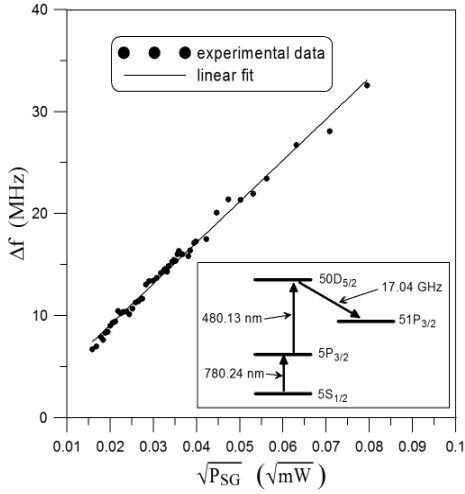}}
\caption{Experimental data for the measurement for $\Delta f$ at 17.04~GHz.}
\label{deltaf17}
\end{figure}

We first perform measurements at 17.04~GHz. For this case the red laser power level at the input to the cell is 175~nW.  A horn antenna is used and connected to a signal generator (SG) via a 4-m cable. The horn antenna is 0.88~m from the center of the two laser beams. During the experiments, the power level on the SG is varied from -10~dBm to 10~dBm (or 0.1~mW to 10~mW).  The blue laser is tuned to $\approx 480.13$~nm to couple states $5{\rm P}_{3/2}$ and $50{\rm D}_{5/2}$, and the 17.04~GHz field couples $50{\rm D}_{5/2}-51{\rm P}_{3/2}$, see Fig.~\ref{deltaf17}. The power of the blue laser is 30~mW.

\begin{figure}
\centering
\scalebox{.3}{\includegraphics*{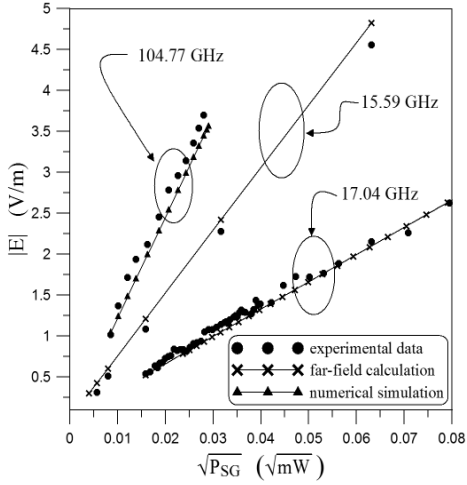}}
\caption{Comparison of experimental data to both numerical simulations and to far-field calculations for 15.59~GHz, 17.04~GHz and 104.77~GHz.}
\label{efield-psg}
\end{figure}

Fig.~\ref{deltaf17} shows the measured $\Delta f$ as a function of the square root of the SG power (labeled as $\sqrt{P_{SG}}$).  We see that the measured $\Delta f$ is linear with respect to $\sqrt{P_{SG}}$ (noting $|E|\propto \sqrt{P_{SG}}$), as predicted from (\ref{mage2}).  With the measured splitting ($\Delta f$), the absolute field strength at the location for the lasers can be obtained with (\ref{mage2}).
Fig.~\ref{efield-psg} shows the calculated E-field strength as a function of $\sqrt{P_{SG}}$. In obtaining the $E$-field values, the required $\wp_{RF}$ was determined from (\ref{wp2}) and the results in Fig.~\ref{rpart}.   As a comparison, we use a far-field calculation for the $E$-field radiating from a horn antenna.  Taking into account the distance, the gain of antenna, and the cable losses, the far-field $E$-field is calculated and the results are also shown in Fig.~\ref{efield-psg}.  What is meant in this comparison is that for each measured $\Delta f$ (corresponding to a given ${P_{SG}}$) we calculate $|E|$ from (\ref{mage2}).  It is this $P_{SG}$ that is used in the far-field calculation.  We see very good agreement between the far-field calculation and the measured $E$-field.

\begin{figure}
\centering
\scalebox{.30}{\includegraphics*{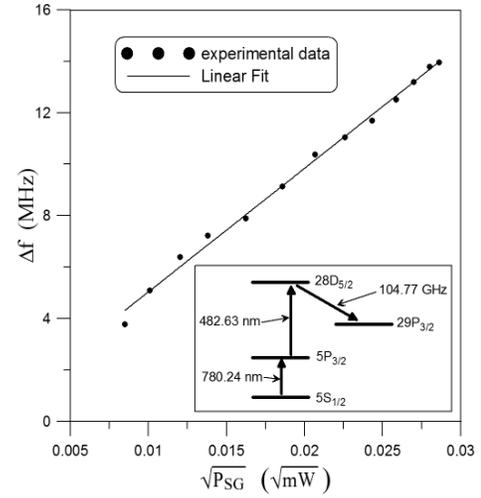}}
\caption{Experimental data for the measurement for $\Delta f$ at 104.77~GHz.}
\label{104data}
\end{figure}

With the same two lasers and vapor cell, it is possible to measure an RF source at much higher frequencies.  To illustrate this, we performed measurements at 104.77~GHz. For this case the red laser power level at the input to the cell is 120~nW. A WR-10 open-ended waveguide is used (see Fig.~\ref{vaporcell}a) as a source antenna and is connected to a signal generator. The waveguide is 139~mm from the laser beams inside the cell. The output power from the waveguide is varied from -11.43~dBm to -0.66~dBm (or 0.072~mW to 0.86~mW). Again we refer to this power as $P_{SG}$.  The blue laser is tuned to $\approx 482.63$~nm to couple states $5{\rm P}_{3/2}$ and $28{\rm D}_{5/2}$, and the 104.77~GHz field couples $28{\rm D}_{5/2}-29{\rm P}_{3/2}$, see Fig.~\ref{104data}. The power of the blue laser is 22~mW.

Fig.~\ref{104data} shows the measured $\Delta f$ as a function of the square root of the waveguide power (labeled as $\sqrt{P_{SG}}$).  We see the measured $\Delta f$ is linear with respect to $\sqrt{P_{SG}}$ as predicted from theory.  Fig.~\ref{efield-psg} shows the calculated E-field strength from (\ref{mage2}) as a function of $\sqrt{P_{SG}}$.  For comparison, we perform a three-dimensional numerical simulation with HFSS (mentioning this numerical code does not imply an endorsement, but serves to clarify the techniques used) in order to determine the $E$-field from the open-ended waveguide at a distance of 139~mm. In the simulation we use the same $P_{SG}$ used in the experiments.  This numerical simulation for $|E|$ is also shown in Fig.~\ref{efield-psg}.  Once again, we see good agreement between the measured and simulated $E$-field, validating this technique.

\begin{figure}[!t]
\centering
\scalebox{.28}{\includegraphics*{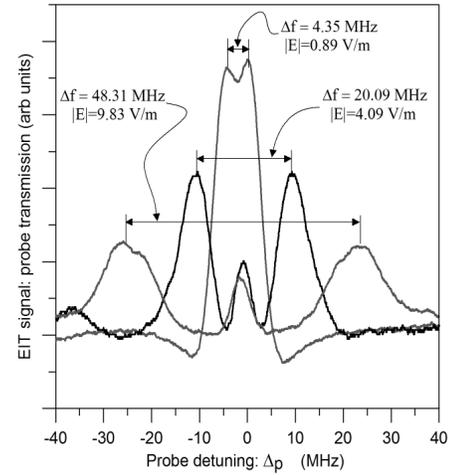}}
\caption{Experimental data for $E$-field measurements at 68.64~GHz (the $32{\rm D}_{5/2}-33{\rm P}_{3/2}$ transition).}
\label{eit68}
\end{figure}

We illustrate another example with the same set-up by showing a measurement at 68.64~GHz. For this case the red laser power level at the input to the cell was 250~nW.  A WR-15 open-ended waveguide is used as a source antenna and is connected to a SG.  The blue laser is tuned to $\approx 481.75$~nm to couple states $5{\rm P}_{3/2}$ and $32{\rm D}_{5/2}$, and the 68.64~GHz field couples $32{\rm D}_{5/2}-33{\rm P}_{3/2}$. The power of the blue laser is 24~mW. Fig.~\ref{eit68} shows the EIT signal for this frequency for the case of RF on and RF off.  We show results for two different $E$-field measurements in the figure, where we see well defined splitting of the EIT-signal. Also shown in this are the calculated $E$-fields, obtained from (\ref{mage2}) and the results of Fig.~\ref{rpart}.

The three examples above were for the $nD_{5/2}\rightarrow(n+1)P_{3/2}$ transition. As discussion above (see Fig.~\ref{freq}), several different frequencies can be measured for these particular transitions. However, additional RF fields strengths can be measured with other atomic transitions (other than $nD_{5/2}\rightarrow(n+1)P_{3/2}$). While we do not discuss these other transitions in detail, we will show two examples for the purpose of illustrating the broadband nature of the probe and for discussing the potential high-power applications.

\begin{figure}[!t]
\centering
\scalebox{.30}{\includegraphics*{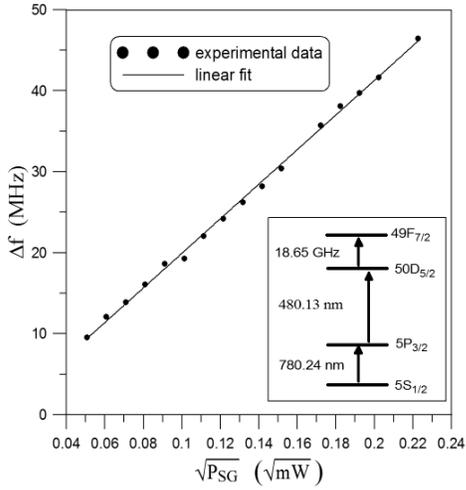}}
\caption{Experimental data for $E$-field measurements at 18.65~GHz.}
\label{deltaf-18}
\end{figure}

In the first example, we use the same setup that is used for the 17.04~GHz measurements above to perform measurements at 18.65~GHz, in that we use the same horn antenna with the blue laser tuned to $\approx 480.13$~nm. This corresponds to $5{\rm S}_{1/2}-5{\rm P}_{3/2}-50{\rm D}_{5/2}$. In this configuration a splitting of the EIT signal can be observed at not only 17.04~GHz (as discussed above), but at 18.65~GHz as well. The 18.65~GHz transition couples $50{\rm D}_{5/2}-49{\rm F}_{7/2}$. Fig.~\ref{deltaf-18} show the measured splitting $\Delta f$ as a function of $\sqrt{P_{SG}}$. Here again we observe a linear relationship between $\Delta f$ and $\sqrt{P_{SG}}$, as predicted by the theory.

\begin{figure}
\centering
\scalebox{.30}{\includegraphics*{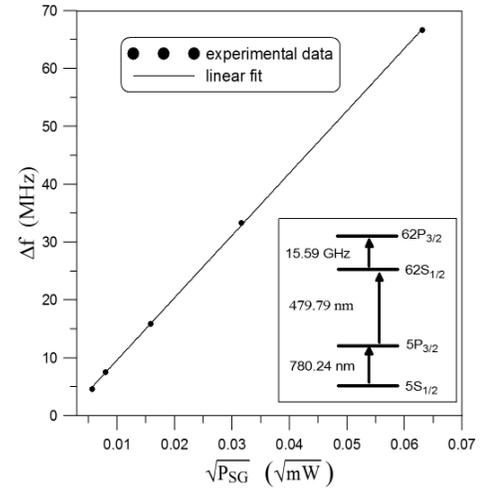}}
\caption{Experimental data for the measurement for $\Delta f$ at 15.59~GHz.}
\label{deltaf15}
\end{figure}

\begin{figure}
\centering
\scalebox{.60}{\includegraphics*{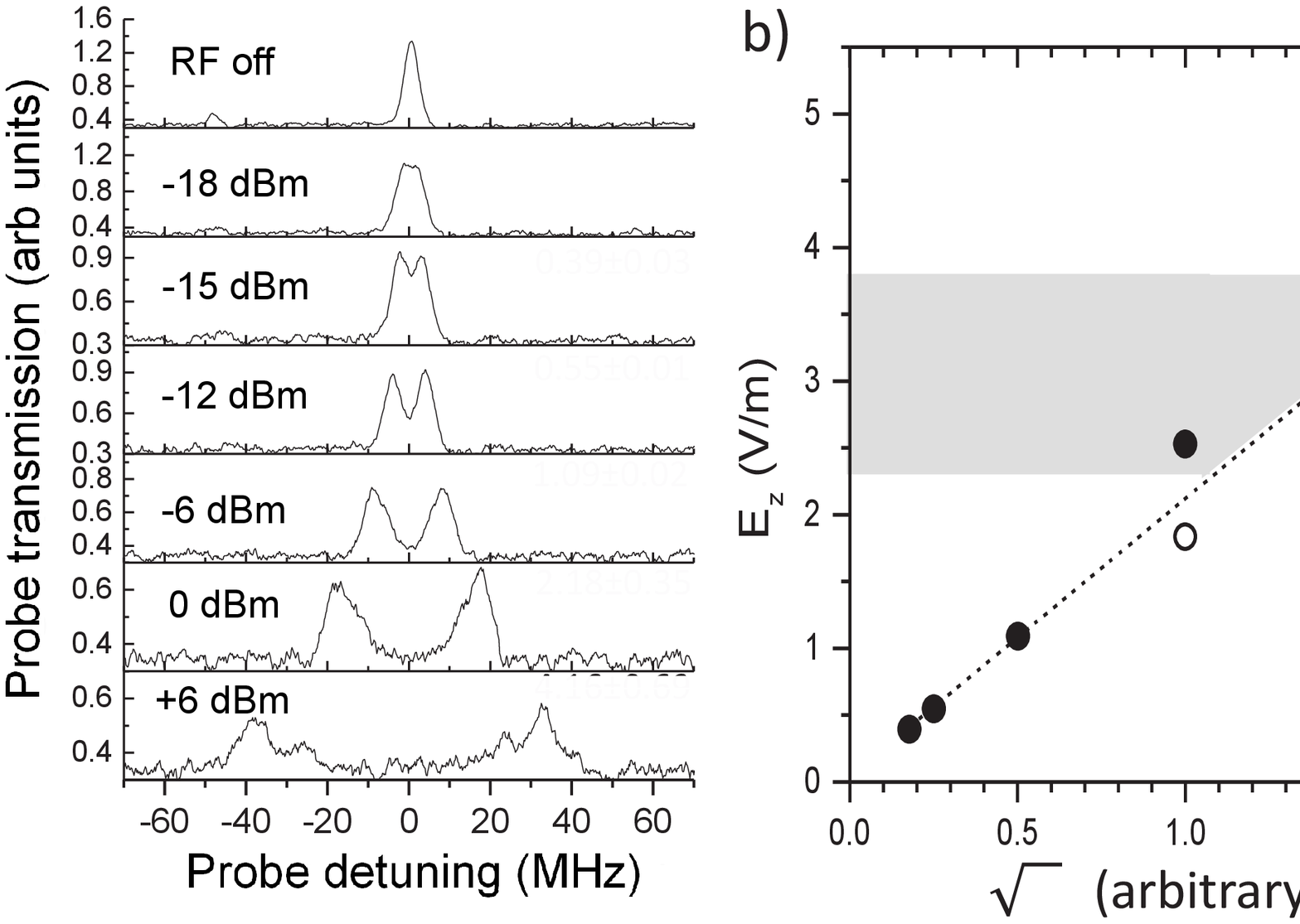}}
\caption{Illustration of splitting and peak reduction for increasing power at 15.59~GHz.}
\label{splot}
\end{figure}

In the last example we show results for a $62S_{1/2}-62P_{3/2}$ transition. In this experiment, we perform measurements at 15.59~GHz.
For this case, the red laser power level at the input to the cell is 220~nW. The blue laser is tuned to $\approx 479.79$~nm to couple states $5{\rm P}_{3/2}$ and $62{\rm S}_{1/2}$, and the 15.59~GHz field couples $62{\rm S}_{1/2}-62{\rm P}_{3/2}$, see Fig.~\ref{deltaf15}. The power of the blue laser is 30~mW. In this setup, the horn antenna is 0.33~m from the center of the two laser beams inside the cell. During the experiments, the power level on the SG is varied from -18~dBm to 6~dBm (or 0.016~mW to 0.39~mW). In Fig.~\ref{splot} we have superimposed the splitting of the EIT signal for different SG power settings.  The figure illustrates the increases in splitting as the field strength at the cell increases. Fig.~\ref{deltaf15} shows $\Delta f$ as a function of $\sqrt{P_{SG}}$. Once again, we see a well defined linear relationship.
Fig.~\ref{efield-psg} shows results for the $E$-field obtained from (\ref{mage2}) as a function of $\sqrt{P_{SG}}$. In this plot, we have also plotted the E-field values obtained from a far-field calculation for these antenna parameters. Once again good agreement is demonstrated between the measured and far-field calculations.

The results in Fig.~\ref{splot} show an interesting point.  From this figure we see that as the field strength is increased, the separation in the splitting not only increases, but equally important, we see that the heights of the two peaks get smaller and become comparable to the back ground noise signal.  Here lies a potential problem.  If the field strength becomes too large, the peaks may be difficult to determine and hence it could be hard to accurately measure these large field strengths.  We are currently looking at methods that will allow the measurement of very high field strengths that will overcome this issue. One approach that has shown promise, is to use the concept of RF ``two-photon'' transitions, see \cite{twophoton} for details.

In this section, we have shown experimental data for the measurement of $E$-field strengths for a wide range of frequencies, and we have shown very good agreement between these measured field strengths to both numerical simulations and far-field calculations.  The results in this section illustrate the broadband nature of this type of measurement technique.


\section{Measurement Uncertainties}

The uncertainties in this type of measurement are still being investigated in detail, but we can comment on some of the aspects.  In general, the uncertainties can be grouped into two different categories: (a) quantum based uncertainties (i.e., parameters and issues related to the atomic physics aspect of the technique) , and (b) RF based uncertainties (i.e., parameters and issues related to the RF aspect of the technique).  Some of the various quantum based uncertainties have been discussed in \cite{jim}.  As seen from (\ref{mage2}), the one calculable parameter needed in this technique is the atomic dipole moment ($\wp_{RF}$). Using the best available quantum defect (\cite{gal2} and \cite{gal3}) to perform a numerical calculation of $\wp_{RF}$ (see discussion in section \ref{dipolesec}) it is believed that $\wp_{RF}$ can be determined to less than $0.1~\%$.

We and other groups are currently looking at methods to obtain more accurate values for $\wp_{RF}$.  With this said, it is believed that the largest source of measurement uncertainties in this technique is the RF based uncertainty.  This type of uncertainty results from the fact that since we are using a glass (or dielectric) cell to hold the atoms, the RF field interacts with the cell itself.  When an RF wave is incident onto a hollow glass cell, standing waves can develop on the inside of the cell due to the internal reflection inside the cell. The distribution of the E-fields inside the cell will vary depending on the frequency and on the size of the cell. This may result in the field inside the cell being different than the incident field (the desired measured quantity).  It is unclear how large of an effect this is, and we are currently investigating this issue. With that noted, this perturbation can be reduced by making the cell as small as possible. If this perturbation is not reduced entirely, it can be calculated and accounted for. We have recently manufactured cells on the order of 2~mm and smaller in size (see next section), and future work will be looking at these cell-size efforts on field perturbation.

\section{Applications}

The obvious applications of the type of technique include: (1) a direct SI unit link RF $E$-field measurement, (2) the technique will be self calibrating: traced to atomic transitions, (3) the technique will provide an RF measurement independent of current techniques, (4) the technique will measure both very weak and very strong fields over a large range of frequencies ($\sim$0.1~mV/m: two to three orders of magnitude improvement over current approaches; and $<$0.01~mV/m may be possible), and (5) the probe/sensor can be used as a stand alone measurement or it can be used to calibrate existing probes and/or test facilities.

Beside the obvious applications, the technique can open up numerous other applications. Some of the other interesting applications include: (1) having a truly broadband probe: covering 1~GHz to 500~GHz, (2) the possibility of a compact in vitro/vivo measurement, e.g., specific absorption rate (SAR) probe, (3) bio-sensing and bio-imaging on a compact scale, (3) imaging sensors and arrays of small vapor cells,  (4) the abilities of field measurements in small confined spaces, (5) millimeter wave and sub-terahertz wave traceable calibrations (not currently possible) \cite{gordon}, and (6) the possibility of measurements on a small spatial scale, i.e., sub-wavelength imaging and field mapping over a large frequency range.

\begin{figure}[!t]
\centering
\scalebox{.4}{\includegraphics*{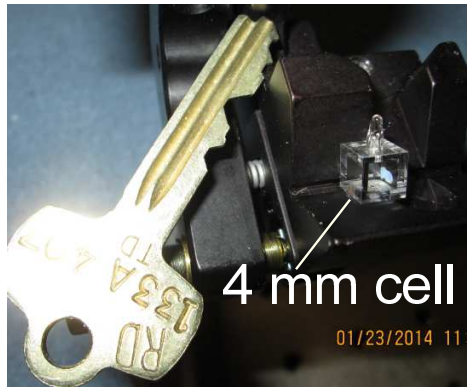}}
\centerline{\footnotesize{(a) 4-mm cell}}
\scalebox{.3}{\includegraphics*{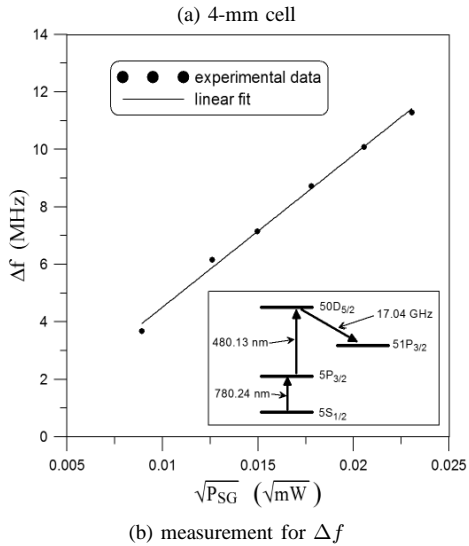}}
\centerline{\footnotesize{(b) measurement for $\Delta f$}}
\caption{4-mm vapor cell experiments.}
\label{4mmcell}
\end{figure}

\begin{figure}[!t]
\centering
\scalebox{1.5}{\includegraphics*{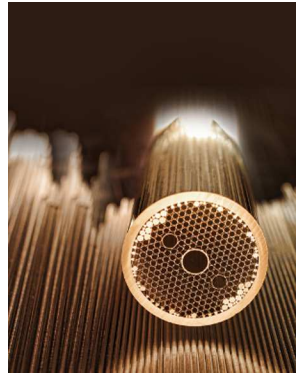}}
\caption{Hollow-core photonic bandgap fiber based vapor cell. Photo courtesy of the DARPA's website.}
\label{fiber}
\end{figure}

Most notable of these listed applications are the sensitivity capability, the compact nature, and sub-wavelength imaging aspects of the technique. While in the experimental examples shown here, we use a cell on the order of 25~mm to 75~mm, the vapor cell can be made smaller and hence allow a compact probe (or sensor head).  We are currently investigating this aspect, where in we are performing experiments on 2~mm and 4~mm cubic cells, see Fig.~\ref{4mmcell}(a).  While this work will be reported in detail later, Fig.~\ref{4mmcell}(b) shows initial data taken with the 4~mm cell. The data show that for this small cell, the desired linear behavior is observed.  Our eventual goal in this work is to use a hollow-core photonic bandgap fiber as a vapor cell (see Fig.~\ref{fiber}), resulting in a probe with a spatial size on the order 100's~$\mu$m and smaller. We are currently investigating such a fiber.

Regardless of the size of the vapor cell, this technique allows for sub-wavelength imaging of an RF field over a large frequency range.  This has been demonstrated in \cite{imaging} where field distributions inside a glass cell were imaged at both 17.04~GHz and 104.77~GHz.  The unique feature of this imaging approach is that the spatial resolution is not governed by the size of the vapor cell that holds the atoms. In fact, the spatial resolution is governed by the width of the laser beams used.  The RF field will only interact with the atoms that are exposed to the two laser beams. As such, the spatial resolution of this approach is based on beam widths of the two lasers used in this experiment, which can be in principle on the order of the diffraction limit, i.e., 10's~$\mu$m.  The applications of such a small spatial imaging capability are numerous. For example, the sensing volume could be scanned over a printed-circuit-board (PCB) or a metasurface \cite{metasurface} in order to map their fields, as well as other applications where E-field measurements on a small spatial resolution are desired.

\section{Discussion and Conclusion}

We have presented the framework for a technique using Rydberg atoms place in a vapor cell for measuring $E$-field strengths at RF.  The benefit of using atoms allows us to convert an amplitude measurement of the RF field strength to a frequency measurement of an optical signal, giving both an accurate and sensitive $E$-field strength measurement.  We have demonstrated that this is a ``truly'' broadband technique. That is, with two tunable lasers and one vapor cell, it is possible to measure RF field strengths over a frequency range of 1~GHz to 500~GHz.  We have validated this technique by comparing experimental data to both numerical simulations and to far-field calculations for various frequencies from 15~GHz to 105~GHz.

This technique has the capability of performing a direct traceable SI measurement that does not need to be calibrated, has drastically improved sensitivity, and can be very compact ($<$1~mm).   Besides the obvious uses, the probe can have numerous new applications. Most notably are the sensitivity and sub-wavelength measurement aspects. From a sensitivity viewpoint, at least two to three orders of magnitude improvement over current techniques are attainable and field strength measurements of $<$0.01~mV/m may be possible. This technique can also be used to perform sub-wavelength imaging and field mapping over a large range of frequencies \cite{imaging}. The E-field imaging volume is determined by the overlap of the RF, the probe beam, and coupling beam within the vapor cell, and as such the high spatial resolution is based on the two lasers' beam widths, which can be on the order of 50~$\mu$m to 100~$\mu$m (and possibly smaller). In future publications we will address a detailed uncertainty analysis, present various imaging measurements and applications, as well as discuss the scaling down of the technique in order to make a compact field probe/sensor.

\section{Acknowledgment} The authors thank Galen H. Koepke and Dr. Thomas Heavner from the National
Institute of Standards and Technology (NIST) for their helpful suggestions and technical discussions.

\end{document}